\documentclass[acmsmall,screen]{acmart}

\def\BibTeX{{\rm B\kern-.05em{\sc i\kern-.025em b}\kern-.08em
    T\kern-.1667em\lower.7ex\hbox{E}\kern-.125emX}}
\usepackage{paralist}
\usepackage{amsmath,amsfonts}
\usepackage{array}
\usepackage{textcomp}
\usepackage{stfloats}
\usepackage{url}
\usepackage{verbatim}
\usepackage{graphicx}
\usepackage[ruled,vlined,linesnumbered]{algorithm2e}
\usepackage{algpseudocode}
\usepackage{xcolor}
\usepackage{multirow}
\usepackage{amsfonts}
\usepackage{xspace}
\usepackage{colortbl}
\usepackage{makecell}
\usepackage{microtype}
\usepackage{enumitem}
\usepackage{pifont}
\usepackage{booktabs}
\usepackage[switch]{lineno}
\usepackage{tcolorbox}
\newtcolorbox{ansbox}{colframe = gray!75!black}
\usepackage{subcaption}

\newcommand{\tool}{\textit{MoDitector}\xspace}
\newcommand{\oracle}{\textit{Module-Specific Oracle}\xspace}
\newcommand{\select}{\textit{Adaptive Seed Generation}\xspace}
\newcommand{\feedback}{\textit{Module-Specific Feedback}\xspace}

\newcommand{\mccs}{$\mathcal{M}\text{ICS}$\xspace}
\newcommand{\abb}{\textit{MoDitector}\xspace}

\title{\tool: Module-Directed Testing for Autonomous Driving Systems}

\author{Renzhi Wang}
\authornote{Both authors contributed equally to this research.}
\authornote{This work was done during the author's visit to Singapore Management University}
\orcid{0000-0002-1219-0732}
\affiliation{%
  \institution{University of Alberta}
  \city{Edmonton}
  \state{Alberta}
  \country{Canada}}
\email{renzhi.wang@ualberta.ca}

\author{Mingfei Cheng}
\authornotemark[1]
\orcid{0000-0002-8982-1483}
\affiliation{%
  \institution{Singapore Management University}
  \city{Singapore}
  \country{Singapore}
}
\email{mfcheng.2022@smu.edu.sg}

\author{Xiaofei Xie}
\orcid{0000-0002-1288-6502}
\affiliation{%
 \institution{Singapore Management University}
 \city{Singapore}
 \country{Singapore}}
\email{xfxie@smu.edu.sg}

\author{Yuan Zhou}
\authornote{Corresponding author.}
\orcid{0000-0002-1583-7570}
\affiliation{%
  \institution{Zhejiang Sci-Tech University}
  \city{Hangzhou}
  \state{Zhejiang}
  \country{China}}
\email{zhouyuan1130@163.com}

\author{Lei Ma}
\orcid{0000-0002-8621-2420}
\affiliation{%
  \institution{The University of Tokyo \& University of Alberta}
  \city{Tykyo}
  \country{Japan}}
\email{ma.lei@acm.org}

\date{\today}

\begin{document}

\begin{abstract}
Testing Autonomous Driving Systems (ADS) is crucial for ensuring their safety, reliability, and performance. Despite numerous testing methods available that can generate diverse and challenging scenarios to uncover potential vulnerabilities, these methods often treat ADS as a black-box, primarily focusing on identifying system failures like collisions or near-misses without pinpointing the specific modules responsible for these failures. Understanding the root causes of failures is essential for effective debugging and subsequent system repair. We observed that existing methods also fall short in generating diverse failures that adequately test the distinct modules of an ADS, such as perception, prediction, planning and control.

To bridge this gap, we introduce \tool, the first root-cause-aware testing method for ADS. Unlike previous approaches, \tool not only generates scenarios leading to collisions but also showing which specific module triggered the failure. This method targets specific modules, creating test scenarios that highlight the weaknesses of these given modules. Specifically, our approach involves designing module-specific oracles to ascertain module failures and employs a module-directed testing strategy that includes module-specific feedback, adaptive seed selection, and mutation. This strategy guides the generation of tests that effectively provoke module-specific failures. We evaluated \tool across four critical ADS modules and four testing scenarios. The results demonstrate that our method can effectively and efficiently generate scenarios where errors in targeted modules are responsible for ADS failures. It generates 216.7 expected scenarios in total, while the best-performing baseline detects only 79.0 scenarios. 
Our approach represents a significant innovation in ADS testing by focusing on identifying and rectifying module-specific errors within the system, moving beyond conventional black-box failure detection.
\end{abstract}

\keywords{Module-Specific Failure, Autonomous Driving System, Testing}
\maketitle

\section{Introduction}\label{sec: Introduction}

Autonomous Driving Systems (ADSs) have rapidly advanced in recent years, aiming to enhance transportation and urban mobility. ADSs act as the brain of autonomous vehicles (AVs), processing extensive sensor data (e.g., images, LiDAR points) through various modules (e.g., perception, prediction, planning, and control) to operate vehicles. ADSs are considered highly safety-critical systems, as they have demonstrated vulnerability to critical issues arising from minor perturbations in the driving environment. For example, Tesla’s Autopilot system has been involved in multiple accidents where it failed to recognize stationary emergency vehicles, leading to crashes and prompting federal investigations~\cite{electrek2023tesla}.
Therefore, ADSs require extensive testing to ensure their safety and reliability during real-world deployment.  

ADS testing can be broadly divided into \textit{real-world testing} and \textit{simulation-based testing}. Real-world testing is necessary, but it is impractical and costly due to the vast number of miles required and the rarity of critical events. In contrast, simulation-based testing offers a controllable, efficient, and cost-effective environment for both developing and testing autonomous vehicles (AVs), allowing for the creation and testing of diverse scenarios that would be difficult to reproduce in real-world conditions. In recent years, significant efforts have been made to advance simulation-based testing for ADSs to identify potential system failures such as collision and timeout. 
Specifically, existing studies have primarily focused on reconstructing scenarios from real-world data~\cite{van2015automated,fang2020augmented,kruber2018unsupervised} or developing search algorithms~\cite{rana2021building,cheng2023behavexplor,zhang2022adversarial} to generate safety-critical scenarios.

An ADS typically comprises multiple modules, including perception, prediction, planning, and control, each tasked with specific functions. Recent research~\cite{lou2022testing} underscores the importance of understanding the root causes of detected failures, which poses challenges in effectively detecting scenarios that reflect diverse root causes. While current methods can identify numerous safety-critical scenarios, they often treat the ADS as a black box. It remains unclear \textit{which specific module leads to a failure and how the diversity of generated failures accurately reflects the weaknesses of different modules}. Existing black-box testing methods may be biased toward detecting failures predominantly caused by weaker modules, such as those in planning or perception, thus not providing a understanding and comprehensive view of all potential weaknesses (see results in Section~\ref{sec: perliminary_study}). Detecting failures across various modules is crucial for developers to understand the root causes accurately and enhance the corresponding components effectively.

One might question whether it is sufficient to test individual modules within an ADS, such as evaluating the object detection model in the perception module. Indeed, some studies~\cite{ma2024slowtrack, zhang2024data} have focused on model-level evaluations, testing specific modules like perception. However, model errors do not always lead to system failures, and the multiple modules within an ADS can often tolerate some model errors (refer to Section~\ref{sec: perliminary_study}). For instance, an error in the perception module to detect a distant vehicle may not necessarily lead to a system failure, such as a collision, particularly if the missed vehicle does not impact the trajectory of the ego vehicle. Therefore, a more effective approach to ADS testing should generate diverse \textit{system-level} failure, which can pinpoint the limitations of individual modules. It is the primary goal of this paper.

To fill this gap, we aim to investigate the generation of ADS system failures—critical scenarios predominantly triggered by specific modules, such as perception, prediction, planning and control. We define these scenarios as 
``\textbf{\underline{M}}odule-\textbf{\underline{I}}nduced \textbf{\underline{C}}ritical \textbf{\underline{S}}cenarios'' (\mccs), which are safety-critical scenarios revealing system-level failures caused by malfunctions within a designated module $\mathcal{M}$. For instance, consider a collision precipitated by a mis-detection in the perception module while other modules operate correctly. Nonetheless, generating \mccs for a particular module $\mathcal{M}$ presents two primary challenges: 
\ding{182} The first challenge is the lack of effective oracles for identifying \mccs. Intuitively, generating \mccs requires the target module to exhibit errors while other modules function correctly. This necessitates an oracle capable of accurately determining whether a module is malfunctioning or operating as intended. For example, the interactive nature of modules in ADSs complicates the attribution of failures, making it difficult to discern whether a failure originates from the planning module or upstream modules like perception and prediction.
\ding{183} Another challenge lies in effectively and efficiently generating {\mccs}s considering the complex interactions of multiple modules. For instance, most existing methods rely on safety-critical feedback, such as minimizing distance to collision, which may not effectively generate \mccs for the given target module, $M$. As noted earlier, the optimization process in testing tools may be biased towards identifying failures induced by the least robust module, rather than the target module $M$. Therefore, it is essential to employ an effective and efficient optimization strategy that specifically generates \mccs caused by errors in the target module, leading to system failures.

To address these challenges, we propose \tool, a Module-Directed Testing framework for Autonomous Driving Systems, designed to automatically generate {\mccs}s that reveal ADS failures for different modules. \tool comprises three main components: \oracle, \feedback and \select. 
To tackle the first challenge \ding{182}, we design various module-specific metrics that can be used as the oracles in the ADS, allowing each module's performance to be evaluated independently. During the testing phase, \oracle can filter failures that are not caused by the errors of the target module. 
To address the second challenge \ding{183}, we introduce \feedback and \select to improve the optimization of detecting {\mccs}s. 
Specifically, \feedback provides feedback that reflects the extent of errors in different modules, enabling \tool to search for {\mccs}s by maximizing errors in the targeted module while minimizing errors in other modules. 
Furthermore, \select enhances testing performance by incorporating adaptive seed selection and mutation strategies. It prioritizes seeds with higher feedback values and adaptively adjusts mutation strategies based on feedback, thereby increasing the performance of detecting {\mccs}s.

We have evaluated \tool on the ADS platform Pylot~\cite{gog2021pylot} using the high-fidelity simulator CARLA~\cite{dosovitskiy2017carla}. 
We compared \tool with three baseline methods: \textit{Random}, i.e., which generates scenarios randomly, along with two state-of-the-art ADS testing methods, i.e., AVFuzzer~\cite{li2020av} and BehAVExplor~\cite{cheng2023behavexplor}. 
The evaluation results demonstrate the effectiveness and efficiency of \tool in detecting {\mccs}s, uncovering ADS failures across four different modules. 
For example, \tool generates a total of 55.3, 75.3, 71.7, and 14.3 {\mccs}s for the perception, prediction, planning, and control modules across four testing scenarios, respectively, while the best-performing baseline detects only 18.3, 25.3, 34.3, and 7.0 {\mccs}s for these modules.
Further experimental results demonstrate the fidelity of the designed \oracle, the effectiveness of \feedback, and the usefulness of the \select component within \tool.

In summary, this paper makes the following contributions:
\begin{enumerate}[leftmargin=*]

\item To the best of our knowledge, this is the first study to investigate the generation of Module-Induced Critical Scenarios, which highlight specific weaknesses within individual modules of the ADS.

\item We conduct an empirical study on existing methods, revealing limitations in module-based testing for causing system failures and in system-level testing for covering weaknesses across diverse modules in the ADS.

\item We have developed a novel testing framework that effectively and efficiently generates {\mccs}s. This framework provides fine-grained evaluation results at the module level. To assess module correctness, we have designed specific metrics that serve as oracles for various modules.

\item We conduct extensive experiments to evaluate the effectiveness and usefulness of \tool on Pylot. On average, 13.8, 18.8, 17.9, and 2.6 {\mccs}s were discovered across four initial scenarios for the four main modules (perception, prediction, planning, and control), demonstrating \tool's value in detecting module-specific failures.
\end{enumerate}

\section{Background}\label{sec: Background}

\subsection{Autonomous Driving System}\label{sec: Background-ads}
Autonomous Driving Systems (ADSs) rely on advanced artificial intelligence algorithms as the brain of autonomous vehicles, governing their movements. ADSs can be broadly categorized into two types: End-to-End (E2E) ADSs and Module-based ADSs.
E2E ADSs, such as Openpilot \cite{openpilot}, approach the system as a single deep learning model that processes sensor data (e.g., images) as input and directly outputs control commands to operate the vehicle. The development of E2E ADSs requires vast amounts of data to achieve satisfactory performance levels, which limits their generalization capabilities and industrial-level reliability.
In contrast, module-based ADSs, such as Pylot \cite{gog2021pylot}, Apollo \cite{baiduapollo}, and Autoware \cite{Autoware}, demonstrate more reliable performance in industrial applications by incorporating multiple modules dedicated to specific tasks. Therefore, in this paper, we focus on testing module-based ADSs to highlight critical issues arising from a specified module. 
Formally, we denote the module-based ADS under test as $\mathcal{A} = \{{M}^{1}, \ldots, {M}^{K}\}$, where ${M}^{{i}}$ represents an individual module within the ADS, and $K$ is the total number of modules in the system.
Typically, with perfect sensor data (e.g., GPS signals) and map data, module-based ADSs consist of four primary functional modules: \textit{Perception}, \textit{Prediction}, \textit{Planning}, and \textit{Control}.

\textit{Perception.} 
The Perception module aims to detect objects (e.g., vehicles, pedestrians) surrounding the ego vehicle by processing sensor data (e.g., images, LiDAR points). Formally, at timestamp $t$, given sensor data $\mathcal{I}_{t}$, the perception module outputs a sequence of detected objects in the format of bounding boxes, represented as $\mathcal{B}_{t} = \{B_{t}^{1}, \ldots, B_{t}^{N_{t}}\}$, where $B_{t}^{i}$ represents the predicted bounding box of the $i$-th object and $N_{t}$ is the number of detected objects.

\textit{Prediction.} The Prediction module aims to anticipate the future movements of detected objects. Using the outputs from the \textit{Perception} module, it predicts the trajectories of these objects over a future time horizon $\Delta t$. Formally, at timestamp $t$, given the detected objects $\mathcal{B}_{t}$, the prediction module outputs a sequence of future trajectories for each object, represented as $\mathcal{T}_{t}  = \{\tau_{t}^{1}, \ldots, \tau_{t}^{|\mathcal{B}_{t}|}\}$, where $\tau_{t}^{i}$ is the predicted trajectory of the $i$-th object, and $|\mathcal{B}_{t}|$ is the number of detected objects. Each trajectory $\tau_{t}^{i} = \{p_{t+k}^{i} \mid k = 1, \ldots, H_{\text{pred}}\}$ includes a series of future waypoints over a specified prediction horizon $H_{\text{pred}}$, indicating the number of future timesteps for which predictions are made.

\textit{Planning.} The \textit{Planning} module is responsible for determining a safe and efficient path for the ego vehicle, considering the current road conditions and the predicted movements of surrounding objects. Based on the outputs from both the \textit{Perception} and \textit{Prediction} modules, it computes a trajectory for the ego vehicle that adheres to safety constraints and follows driving goals, such as reaching a destination or maintaining lane position. Formally, at timestamp $t$, given the detected objects $\mathcal{B}_{t}$ and predicted trajectories $\mathcal{T}_{t}$, the planning module generates a planned path $\mathcal{P}_{t}$ for the ego vehicle. Here, $\mathcal{P}_{t} = \{p_{t+k}^{\mathcal{A}} \mid k = 1, \ldots, H_{\text{plan}}\}$ is a sequence of waypoints that the ego vehicle should follow over a specified planning horizon $H_{\text{plan}}$.

\textit{Control.} The \textit{Control} module translates the planned path from the \textit{Planning} module into actionable control commands that adjust the ego vehicle’s steering and acceleration to follow the planned trajectory. This module works at a high frequency to ensure real-time responsiveness and vehicle stability. At timestamp $t$, given the planned path $\mathcal{P}_{t}$, the control module generates a control command, represented as $\mathcal{C}_{t} = \{C_{t}^{\text{steer}}, C_{t}^{\text{throttle}}, C_{t}^{\text{brake}}\}$, where $C_{t}^{\text{steer}}$, $C_{t}^{\text{throttle}}$, and $C_{t}^{\text{brake}}$ represent the steering, throttle, and braking commands to be executed by the vehicle at timestamp $t$, respectively.

\subsection{Scenario}
\subsubsection{Scenario Definition} In ADS testing, scenarios represent driving environments structured as sequences of scenes, each capturing a snapshot of scenery and dynamic objects. Scenarios consist of configurable static attributes (e.g., map, weather) and dynamic attributes (e.g., behaviors of NPC vehicles), derived from Operational Design Domains (ODDs)~\cite{thorn2018framework}.
Given the vast attribute space, it is impractical to test all attribute combinations. Following previous studies~\cite{cheng2023behavexplor, li2020av}, we focus on subsets aligned with specific testing goals. In this study, our objective was to test the safety of a module-based ADS, including perception, prediction, planning, and control modules, by configuring the trajectories and weather parameters of NPC vehicles.
Formally, a \textit{scenario} can be defined as a tuple $s = \{\mathbb{A}, \mathbb{P}, \mathbb{E}\}$, where $\mathbb{A}$ represents the ADS motion task, including the start position and destination; $\mathbb{P}$ is a finite set of objects, encompassing static obstacles, dynamic NPC vehicles, and pedestrians; and $\mathbb{E}$ is the set of weather parameters, such as cloudiness. For each object $P \in \mathbb{P}$, we represent its behavior with a sequence of waypoints, denoted by $W^{P} = \{w_{1}, \ldots, w_{n}\}$. Each waypoint $w_{i}$ specifies the location $(x_{i}, y_{i})$ and velocity $v_{i}$. 

\subsubsection{Scenario Observation} A scenario observation is a sequence of scenes recorded during the execution of the scenario, where each scene represents the states of the ego vehicle and other participants at a specific timestamp. Formally, given a scenario $s$, its observation is denoted as $\mathcal{O}(s) = \{o_0, o_1, \ldots, o_{T}\}$, where $T$ is the length of the observation, and $o_t = \{\mathcal{A}_{t}, \mathcal{Y}_t\}$ is the scene at timestamp $t$, including the observation $\mathcal{A}_{t}$ extracted from the ADS and $\mathcal{Y}_t$ from the simulator. In detail, $\mathcal{Y}_t = \{ y_t^\mathcal{A}, y_t^{P_{1}}, \ldots, y_t^{P_{|\mathbb{P}|}}\}$ contains the states of all objects, including the ego vehicle ($y_t^\mathcal{A}$), where $y_t^P = \{p_{t}^{P}, b_{t}^{P}, \theta_{t}^{P}, v_{t}^{P}, a_{t}^{P}\}$ denotes the waypoint of participant $P \in \mathbb{P}$ at timestamp $t$. Each waypoint includes the center position $p_{t}^{P}$, bounding box $b_{t}^{P}$, heading $\theta_{t}^{P}$, velocity $v_{t}^{P}$, and acceleration $a_{t}^{P}$.
Note that the ADS observation $\mathcal{A}_{t}$ contains the outputs of all ADS modules at timestamp $t$.

In the following content, to better distinguish between observations from the ADS and the simulator, we use $\mathcal{A}(s)$ and $\mathcal{Y}(s)$ to denote ADS observation and Simulator observation, respectively. 

\section{Problem Definition and Motivation}

\subsection{Problem Definition}\label{def:mccs}
This paper aims to detect Module-Induced Critical Scenarios ({\mccs}s) for a specified module, defined as follows:

\begin{definition} [$\mathcal{M}$-Induced Critical Scenario] \label{def-mccs} 
Given a ADS $\mathcal{A} = \{{M}^{1}, \ldots, {M}^{K}\}$ that considers multiple modules as well as a target module $\mathcal{A}\in \mathcal{A}$ to be tested, the $\mathcal{M}$-Induced Critical Scenario $s$ satisfies the conditions: 
\begin{itemize}
    \item[a.] $s \in \mathbb{S}^{Fail}$ 
    \item[b.] $\exists s_i\in s. \ error(s_i, \mathcal{M})=True$
    \item[c.] $\forall s_i \in s. \ \forall M\in \mathcal{A}. \ M\neq \mathcal{M} \wedge error(s_i, M)=False$
\end{itemize}
where $\mathbb{S}^{Fail}$ denotes the set of scenarios containing ADS failures, $s_{i}$ is a scene in $s$ ,and the $error$ function determines whether the module $\mathcal{M}$ exhibits errors in a specific scene. Intuitively, if we identify a critical scenario $s$ that results in a failure, and only $\mathcal{M}$ induces errors while all other modules operate correctly across all scenes in $s$, then we can conclude that the failure is primarily caused by $\mathcal{M}$.
\end{definition}

Note that while failures that do not meet the \mccs conditions may still be useful, they do not align with our objectives as we aim to evaluate the quality of individual modules within the ADS. Specifically, we need to accurately localize the root cause in terms of specific modules. If several modules exhibit errors in a critical scenario, it becomes challenging to conclusively determine which module is the root cause. Hence, it is not a good case for developers to analyze and repair. Furthermore, based on our definition, this situation highlights our two main challenges: 1) the $error$ function, which determines whether a module functions correctly, and 2) the effective method to identify \mccs $s$ that satisfies all necessary conditions.

\subsection{Preliminary Study}\label{sec: perliminary_study}

Based on the problem definition, we would like to understand the limitations of existing methods in detecting \mccs. Specifically, both end-to-end system-level testing and module-level testing may generate failures that reveal limitations of individual modules. Therefore, we first conduct an empirical study to evaluate: 1) whether failures generated by system-level testing adequately reflect the diversity of module weaknesses, and 2) whether errors identified through module-level testing can trigger system failures.

\subsubsection{The ability of existing scenario-based testing methods to generate \mccs}\label{sec:perliminary_exist_mccs}

\begin{table}[]
    \centering
    \caption{Module Failures and Collision Distributions of Exising Methods}
    \vspace{-10pt}
    \resizebox{0.65\linewidth}{!}{
    \begin{tabular}{c|ccccc}
    \toprule
         Method & $\mathcal{M}^{\text{Perc}}$ICS & $\mathcal{M}^{\text{Pred}}$ICS & $\mathcal{M}^{\text{Plan}}$ICS & $\mathcal{M}^{\text{Ctrl}}$ICS & Non-\mccs\\
         \midrule
         AVFuzzer & 9  & 17  & 23   & 2 & 47\\
         BehAVExplor & 6 & 57 & 9&  3 & 190 \\
         \bottomrule
    \end{tabular}}
    \vspace{-10pt}
    \label{tab: preliminary_module}
\end{table}

Existing scenario-based methods aim to identify test scenarios that cause ADS failures efficiently but lack root cause analysis for module errors. To explore the capabilities of existing methods in generating \mccs, we conducted experiments using two scenario-based testing scenario generation methods, AVFuzzer\cite{li2020av} and BehAVExplor\cite{cheng2023behavexplor}. In these experiments, we utilized Pylot\cite{gog2021pylot} as the tested ADS and CARLA as the simulator. Starting with four basic scenarios (detailed in Section~\ref{sec: Evaluation}), each method was run for 6 hours, and we recorded the module errors and \mccs collected during this period.

The results shown in table~\ref{tab: preliminary_module} of these two existing works show high similarity. They generate many collisions, however, most are non-\mccs, in which multiple modules typically experience errors before a collision occurs. On the other hand, though some \mccs have been generated, the distribution is highly uneven, with most \mccs introduced by the prediction module, while \mccs from other modules are rare. This imbalance makes it challenging for developers to improve the corresponding modules effectively.

\subsubsection{Limitation on module-level evaluation for ADS testing}
\begin{table}[!t]
    \centering
    \caption{The ratio of module errors that can cause system failures}
    \vspace{-10pt}
    \resizebox{0.65\linewidth}{!}{
    \begin{tabular}{c|ccc|ccc|ccc}
    \toprule
     \multirow{2.5}*{Module}   & \multicolumn{3}{c|}{Perception} & \multicolumn{3}{c|}{Prediction} & \multicolumn{3}{c}{Planning} \\
     \cmidrule(lr){2-4}\cmidrule(lr){5-7}\cmidrule(lr){8-10}
      & 10\% & 20\% & 50\% & 0.1m & 0.5m & 1m & 0.1m & 0.2m & 0.5m\\
     \midrule
     R1 & 0.02 & 0.03 & 0.29 & 0.02 & 0.20 & 0.57 & 0.03 & 0.19 & 0.28\\
     R2 & 0.01 & 0.01 & 0.09 & 0.01 & 0.15 & 0.40 & 0.03 & 0.14 & 0.30\\
     R3 & 0.07 & 0.09 & 0.18 & 0.01 & 0.15 & 0.39 & 0.04 & 0.11 & 0.34\\
     \midrule
     Average & 0.03 & 0.06 & 0.19 & 0.01 & 0.17 &0.45 & 0.03 & 0.15 & 0.31\\
    \bottomrule
    \end{tabular}
    }
    \vspace{-10pt}
    \label{tab:preliminary_fail}
\end{table}

To better investigate the relationship between module-level errors and system-level failures in ADS, we manually introduced random noise to the output of the perception, prediction and planning module since the results in table~\ref{tab: preliminary_module} tend to be error-prone. For the perception module, we applied one of three operations—\textit{Zoom In}, \textit{Zoom Out}, and \textit{Random Offset}—randomly to each bounding box. For the prediction and planning modules, we added random perturbations to trajectory nodes. For each module, we established three levels of random error limits: conventional, moderate, and extreme. The specific settings are as follows:
\begin{inparaitem}
    \item Perception: [10\%, 20\%, 50\%];
    \item Prediction: [0.1m, 0.5m, 1m];
    \item Planning: [0.1m, 0.2m, 0.5m].
\end{inparaitem}
We randomly selected 100 normal running scenarios from \ref{sec:perliminary_exist_mccs}, with each experiment only perturbing one module's output. To mitigate the effects of randomness, each experiment was repeated three times.

Table~\ref{tab:preliminary_fail} shows the results of operations after introducing manual injections.
As the results show, aside from experiments with extreme perturbations (the third column for each module), the module errors alone does not effectively lead to system-level failures. With conventional-level perturbations to module outputs, only a few running failures occurred; even with moderate-level perturbations, the failure rate reached only up to 17\%. This suggests a significant gap between module-level errors and system failures, indicating the need for a mapping method to rapidly identify the corresponding \mccs and bridge this gap effectively.

\begin{ansbox}
   \textbf{Motivation:} Existing system-level and module-level testing methods failed to generate {\mccs}s. This limitation motivates us to develop an effective approach to detecting system failures induced by specified modules. 
\end{ansbox}

\begin{figure*}[!t]
    \centering
    \includegraphics[width=0.85\linewidth]{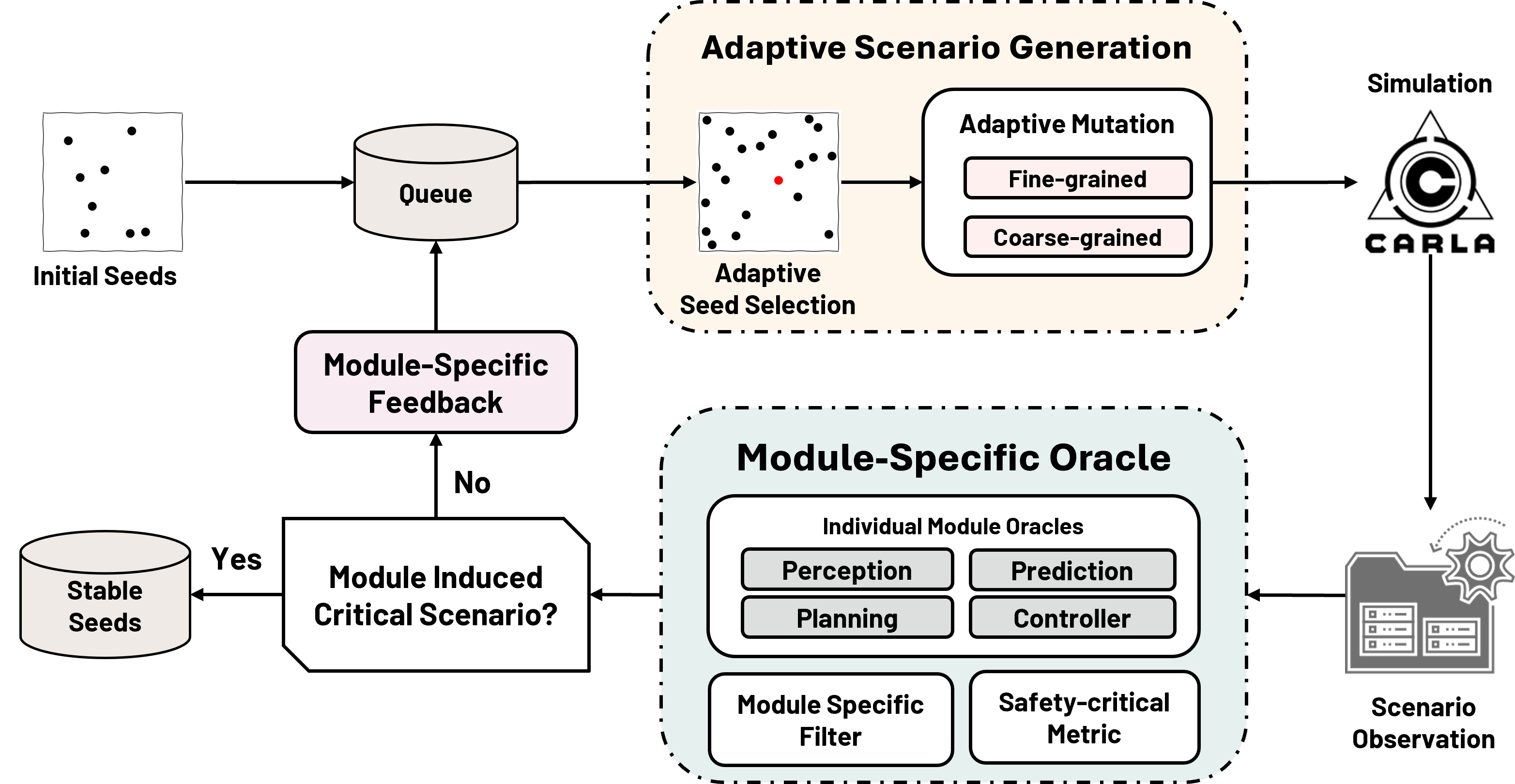}
    \caption{Overview of \tool}
    \label{fig:overview}
\end{figure*}

\section{Approach}\label{sec:method}

\subsection{Overview}
Fig.~\ref{fig:overview} provides a high-level overview of \tool for generating {\mccs}s given initial seeds and the user-specified module $\mathcal{M}$. 
\tool comprises three main components: \oracle, \feedback and \select. 
\oracle functions as an oracle to check whether a scenario satisfies \textit{Definition}~\ref{def-mccs} and qualifies as an \mccs. 
\feedback provides feedback that guides the search process, which jointly considers the system-level specifications (i.e., safety) and the module-specific aspects (i.e., the extent of errors in $\mathcal{M}$).
\select implements an adaptive strategy, including seed selection and mutation, to generate new scenarios based on the module-specific feedback score, thereby improving search performance. 
Specifically, following a classical search-based fuzzing approach, \tool maintains a seed corpus with valuable seeds that facilitate the identification of {\mccs}s. In each iteration, \select first chooses a seed with a higher feedback score and applies an adaptive mutation to the selected seed to generate a new test scenario. Note that a higher feedback score indicates a greater likelihood of evolving into an \mccs. 
Then, \oracle and \feedback provide the identification result and feedback score for the new seed by analyzing the scenario observation, respectively.

Algorithm~\ref{algo:workflow} presents the main algorithm of \tool. The algorithm takes as input an initial seed corpus \( \mathbf{Q} \), a module-based ADS \( \mathcal{A} = \{{M}^{1}, \dots, {M}^{K}\} \), which consists of \( K \) modules, and the user-specified module $\mathcal{M}$ under test.
The output is a set of module $\mathcal{M}$ caused critical scenarios $s$ (Line 13).
In detail, \tool begins by initializing an empty set for $\mathbf{F}_{\mathcal{M}}$ (Line 1). 
Then \tool starts the fuzzing process, which continues until the given budget expires (lines 2-12). 
In each iteration, \tool first uses \select to return a new scenario \( s' \) and the feedback score \(\phi_s\) of its source seed \( s \) (Line 3). 
This new scenario \( s' \) is executed in the simulator with the ADS under test \( \mathcal{A} \), collecting the scenario observation \( \mathcal{O}(s') = \{\mathcal{A}(s'), \mathcal{Y}(s')\} \) including the ADS observation \( \mathcal{O}_{\mathcal{A}}(s') \) and the Simulator observation \( \mathcal{Y}_{\mathcal{A}}(s') \) (Line 4).
Based on these observations, \oracle identifies if the scenario \( s' \) contains system failures and if module \( \mathcal{M}^k \) is the root cause, returning the identification result \( r_{\mathcal{M}} \), module errors $\delta^{\mathcal{A}}$ and safety-critical distance $\delta^{\mathcal{A}}$ (Line 5). 
If \( r_{\mathcal{M}} \) is identified as \textit{Fail}, \tool keeps the scenario \( s' \) in the critical scenario set \(\mathbf{F}_{\mathcal{M}}\) (Line 6-7). 
Otherwise, \feedback calculates a feedback score for the benign scenario \( s' \) based on the module errors \( \delta^{\mathcal{A}} \) and the safety-critical distance \( \delta^{\text{safe}} \) (Lines 8-9).
A higher feedback score \( \phi_{s'} \) indicates a higher potential of \( s' \) for generating {\mccs}s. If the feedback score of seed \( s' \) is higher than that of its parent seed \( s \), \tool retains \( s' \) in the corpus $\mathbf{Q}$ for further optimization (Line 10-11). 
Finally, the algorithm ends by returning $\mathbf{F}_{\mathcal{M}}$ (Line 13).

\begin{algorithm}[!t]
\small
\SetKwInOut{Input}{Input}
\SetKwInOut{Output}{Output}
\SetKwInOut{Para}{Parameters}
\SetKwProg{Fn}{Function}{:}{}
\SetKwFunction{AE}{\textbf{AdaptiveScenariGeneration}}
\SetKwFunction{OI}{\textbf{ModuleSpecificOracle}}
\SetKwFunction{FD}{\textbf{ModuleSpecificFeedback}}
\SetKwComment{Comment}{\color{blue}// }{}
\Input{
Initial seed corpus $\mathbf{Q}$ \\
ADS under test $\mathcal{A} = \{{M}^{1}, ..., {M}^{K}\}$\\
Specified module $\mathcal{M} \in \mathcal{A}$
}
\Output{
$\mathcal{M}$ module-induced critical scenarios $\mathbf{F}_{\mathcal{M}}$
}
$\mathbf{F}_{\mathcal{M}} \gets \{\}$ \\
\Repeat{given time budget expires}{
$s', {\phi}_{s} \gets \AE(\mathbf{Q})$ \Comment{Generate new scenarios}
$\mathcal{O}(s') = \{\mathcal{A}(s'), \mathcal{Y}(s')\} \gets \textbf{Simulator}(s', \mathcal{A})$ \\
$r_{\mathcal{M}}, \delta^{\mathcal{A}}, \delta^{\text{safe}} \gets \OI(\mathcal{A}(s'), \mathcal{Y}(s'), \mathcal{M}^k)$ \Comment{Analyze module errors}
\eIf{$r_{\mathcal{M}}$ is \textit{Fail}}{
    \Comment{Update failure sets}
    $\mathbf{F}_{\mathcal{M}} \gets \mathbf{F}_{\mathcal{M}} \cup \{s'\}$ \Comment{Update discovered {\mccs}s}
}{
    \Comment{Update seed corpus}
    ${\phi}_{s'} \gets \FD(\delta^{\text{safe}}, \delta^{\mathcal{A}})$ \Comment{Calculate feedback score}
    \If{${\phi}_{s'} > {\phi}_{s}$}{
        $\mathbf{Q} \gets \mathbf{Q} \cup \{s'\}$ \Comment{Update corpus for {\mccs}s search}
    }
}
}
\Return $\mathbf{F}_{\mathcal{M}}$
\caption{Workflow of \tool}
\label{algo:workflow}
\end{algorithm}

\subsection{Module-Specific Oracle}
The purpose of \oracle is to serve as an oracle for automatically detecting {\mccs}s by determining whether a given scenario satisfies all conditions outlined in \textit{Definition~\ref{def:mccs}}. This involves two parts: (1) detecting system failures in the scenario (\textit{Definition~\ref{def-mccs}.a}) and (2) determining errors for each module in the ADS (\textit{Definition~\ref{def-mccs}.b} and \textit{Definition~\ref{def-mccs}.c}).

For part (1), we consider the occurrence of collisions as a safety-critical metric to identify system failures.
For part (2), the main challenge is obtaining ground truth to evaluate the performance of individual modules without human annotation. To address this, we design \textit{Individual Module Metrics} to independently measure errors for each module using collected scenario observations, covering the four main modules in the ADS: perception, prediction, planning, and control.

\subsubsection{Safety-critical Metric}\label{sec:safe-metric} We check if the scenario contains ADS failures by detecting collisions. Specifically, we first calculate the minimum distance between the ego vehicle and other objects:
\begin{equation}\label{eq:safe}
    \delta_{s}^{\text{safe}} = \min \left\{ \| R_{\text{bbox}}({p}^{0}_{t}) - R_{\text{bbox}}({p}^{n}_{t}) \|_2 \ \big| \ t \in [0, T], \ n \neq 0 \right\}
\end{equation}
where \( {p}^{0}_{t} \) represents the position of the ego vehicle at time \( t \), \( R_{\text{bbox}}(\cdot) \) calculates the bounding box for an object based on its position \( p \), and \( {p}^{n}_{t} \) represents the position of the \( n \)-th object at the same time. 
Therefore, safety-critical failures can be detected if \( \delta_{s}^{\text{safe}} = 0 \).

\subsubsection{Individual Module Oracles}\label{sec:module-metric}
Given a scenario \( s \) with its observation \(\mathcal{O}(s) = \{\mathcal{A}(s), \mathcal{Y}(s)\}\), we first design individual metrics for each module to measure module-level errors \(\delta^{{M}} = \{\delta^{{M}}_t \ \big| \ t \in [0, \dots, T] \}\) for each module \({M} \in \mathcal{A} \), where \(\delta^{{M}}_t\) denotes the module error at timestamp \( t \) and \( T \) is the termination timestamp of the scenario \( s \).
We detail the calculation of \(\delta^{{M}}_t\) covering the \textit{Perception}, \textit{Prediction}, \textit{Planning}, and \textit{Control} modules as follows:

\noindent \textit{(1) Perception Module.} Given a scenario $s$, the error in the \textit{Perception} module can be directly measured by comparing object bounding boxes between Simulator observation $\mathcal{Y}(s)$ and ADS observations $\mathcal{Y}(s)$.
We adopt a weighted Intersection over Union (IoU)~\cite{girshick2014rich} to measure the errors in the perception module $\mathcal{M}^{\text{perc}}$, which is a widely recognized metric in object detection.
Specifically, the errors of the perception module at timestamp $t$ is calculated as follows:
\begin{equation}
    {\delta^{\text{perc}}_{t}} = 1 - \frac{1}{N_{t}} \sum_{n=1}^{N_{t}} (\frac{D-d_t^{n}}{D} \cdot \frac{|B_t^{n} \cap b_t^n|}{|B_t^{n} \cup b_t^{n}|})
\end{equation}
where \( N_{t} \) represents the number of detected objects within the perception range \( D \) meters, \( B_{t}^{n} \) and \( b_{t}^{n} \) denote the detected bounding box and the ground truth bounding box of the \( n \)-th object, respectively, obtained from the ADS observation $\mathcal{A}_{t} \in \mathcal{A}(s)$ and the scenario observation $\mathcal{Y}_{t} \in \mathcal{Y}(s)$. 
The weight \( \frac{D-d_t^n}{D} \) assigns a higher weight to objects closer to the ego vehicle, where \( d_t^n \) denotes the distance between the \( n \)-th object and the ego vehicle at timestamp \( t \).
A higher value of \(\delta^{\text{perc}}\) represents a greater detection error in the perception module, indicating a potential safety-critical situation.

\noindent \textit{(2) Prediction Module.} Unlike the \textit{Perception} module, directly comparing the predicted trajectories in ADS observation \(\mathcal{A}(s)\) with the collected trajectories in Simulator observation \(\mathcal{Y}(s)\) cannot accurately reflect errors in the \textit{Perception} module. This is because the inputs to the \textit{Prediction} module are derived from the \textit{Perception} module, which may introduce perception errors that subsequently affect prediction outcomes. 
To address this, we adopt a perception-biased trajectory to measure the errors in the \textit{Prediction} module. 
Specifically, at timestamp \( t \), the perception-biased trajectory for the $n$-th object within perception range $D$ is determined by incorporating the biases present in the perception module, formulated as:
\begin{equation}
    \overline{\tau}_t^n = \left\{\ p_{t+k}^n + \Delta p_t^n \mid k \in [0, \ldots, H_{\text{pred}}] \right\}
\end{equation}
where \( p_{t+k}^n \) represents the ground-truth position in the Simulator observation \(\mathcal{Y}_t\), and \( \Delta p_{t+k}^n \) is the position shift observed in the \textit{Perception} module, and $H_\text{pred}$ is the prediction horizon. Note that the position shift \(\Delta p_t^n = (\Delta x_t^n, \Delta y_t^n)\) represents the positional difference between the detected output and the ground truth at timestamp $t$. Since the following detect output after timestamp $t$ is unavailable, and the offset does not fluctuate significantly within a shorter prediction window(about 0.5 to 1 seconds) in most cases, we apply $\Delta p_t^n$ to its predicted sequence.

Consequently, we can measure the error of the \textit{Prediction} module by comparing the predicted trajectories with the perception-biased trajectories. This comparison is quantified by:
\begin{equation}
    \delta_{t}^{\text{pred}} = \max \left\{  \frac{D-d_t^{n}}{D} \cdot \|\hat{p}^{n}_{t+k} - \overline{p}^{n}_{t+k} \|_{2} \ \big| \ \hat{p}^{n}_{t+k} \in \tau_{t}^{n}, \ \overline{p}^{n}_{t+k} \in \overline{\tau}_{t}^{n}, \ n \in [1, \dots, N_{t}] \right\}
\end{equation}
where \( N_{t} \) is the number of detected objects within the perception range \( D \), \( \tau_{t}^{n} \) is the predicted trajectory for the \( n \)-th object, and \( d_t^n \) denotes the distance between the \( n \)-th object and the ego vehicle at timestamp \( t \). The calculation of \(\delta_{t}^{\text{pred}}\) selects the maximum error in all predicted trajectories because this emphasizes the worst-case performance within a given scenario, which is crucial for assessing the robustness and safety of the \textit{Prediction} module. 

\noindent \textit{(3) Planning Module.} We measure errors in the \textit{Planning} module from a safety perspective by evaluating the distance to collisions.
Similar to the prediction module, biases present in upstream modules (i.e., \textit{Perception} and \textit{Prediction}) affect the evaluation of the planning module when directly using ground truth data collected from the Simulator observation \(\mathcal{Y}(s)\). 
To mitigate these biases, we measure errors in the \textit{Planning} module by assessing if the planned trajectories collide with objects detected by the upstream modules. This is calculated by:
\begin{equation}\label{eq:plan}
    \delta^{\text{plan}}_{t} = \sum_{n=1}^{N_{t}} \sum_{k=1}^{H_{\text{plan}}} \mathbb{I}(\| R_{\text{bbox}}(p^{\mathcal{A}}_{t+k}) - R_{\text{bbox}}(p_{t+k}^{n}) \|_{2}=0)
\end{equation}
where \( N_{t} \) represents the number of detected objects, \( H_{\text{plan}} \) is the planning horizon, \( R_{\text{bbox}}(\cdot) \) calculates the bounding box for an object based on its position \( p \), \( \mathbb{I}(\cdot) \) is an indicator function. The indicator function \( \mathbb{I}(condition) \) returns 1 if the \( condition \) is true and 0 otherwise. Additionally, \( p^{\mathcal{A}}_{t+k} \in \mathcal{P}_{t} \) denotes a trajectory point planned by the \textit{Planning} module, and \( p_{t+k}^{n} \in \tau_{t}^{n} \) is the predicted trajectory point for the \( n \)-th object at timestamp \( t \). Ideally, the planned trajectory should be collision-free, maintaining a safety distance from all predicted states of all objects (i.e., \(\delta^{\text{plan}}_{t} = 0\)). Therefore, a larger \(\delta^{\text{plan}}_{t}\) indicates that the planning module has safety-critical errors, such as collisions.

\noindent \textit{(4) Control Module.} 
The control command is directly applied to the vehicle to manage its movement by following a trajectory from the upstream \textit{Planning} module. Obtaining a ground truth for control commands is challenging. Thus, we evaluate the \textit{Control} module by comparing the actual movement of the vehicle with the planned movement from the \textit{Planning} module.
At timestamp \( t \), we calculate the error for the \textit{Control} module by:
\begin{equation}
    \delta^{\text{ctrl}}_{t} = \| p_{t+1}^{\mathcal{A}} - p_{t+1}^{0} \|_{2} + \| v_{t+1}^{\mathcal{A}} - v_{t+1}^{0} \|_{2}
\end{equation}
where \( p_{t+1}^{\mathcal{A}} \) and \( v_{t+1}^{\mathcal{A}} \) represent the planned position and velocity from the \textit{Planning} module at timestamp \( t \), and \( p_{t+1}^{0} \) and \( v_{t+1}^{0} \) are the actual position and velocity of the vehicle at timestamp \( t+1 \). 
This error \(\delta^{\text{ctrl}}_{t}\) quantifies how well the \textit{Control} module is executing the planned trajectory. A larger error value indicates a significant deviation from the planned path and speed, suggesting potential issues in the \textit{Control} module (i.e., inaccuracies in executing the planned trajectory).

\begin{algorithm}[!t]
\small
\SetKwInOut{Input}{Input}
\SetKwInOut{Output}{Output}
\SetKwInOut{Para}{Parameters}
\SetKwProg{Fn}{Function}{:}{}
\SetKwComment{Comment}{\color{blue}// }{}
\Input{
Scenario observation $\mathcal{O}(s) = \{{\mathcal{A}}(s), \mathcal{Y}(s)\}$ \\
ADS $\mathcal{A} = \{{M}^{1}, \dots, {M}^{K} \}$ \\
User-specified module $\mathcal{M}$
}
\Output{
\mccs identification result $r_{\mathcal{M}}$ \\
Module errors $\delta^{\mathcal{A}} = \{\delta^{{M}^{{1}}}, \dots, \delta^{{M}^{{K}}}\}$ \\
Safety-critical distance $\delta^{\text{safe}}$
}

 $\delta^{\mathcal{A}} \gets \{\}$, $r_{\mathcal{M}} \gets Pass$ \\
$\delta^{\text{safe}} \gets \text{calculate safety-critical distance by Eq.~\ref{eq:safe}}(\mathcal{Y}(s))$ \Comment{Safety-critical metric} 
\If{$\delta^{\text{safe}} \neq 0$}{
    $r_{\mathcal{M}} \gets Fail$ \Comment{Fail to satisfy Definition~\ref{def:mccs}.a}
}
\For{${M} \ in \ \mathcal{A}$}{
\Comment{Module error calculated by Individual Module Metrics}
    $\delta^{{M}} \gets \text{calculate the module-level error according to Section~\ref{sec:module-metric}}$ \\
    $\hat{\delta}^{{M}} \gets \text{filter module errors by Eq.~\ref{eq:system_error}}$ \\
    $\delta^{\mathcal{A}} \gets \delta^{\mathcal{A}} \cup \{\delta^{{M}}\}$ \\
    \Comment{Definition~\ref{def:mccs}.b}
    \If{${M} = \mathcal{M} \ and \ \hat{\delta}^{\mathcal{M}} = 0 $}{
        $r_{\mathcal{M}} \gets Fail$ 
    }

    \Comment{Definition~\ref{def:mccs}.c}
    \If{${M} \neq \mathcal{M} \ and \ \hat{\delta}^{{M}} \neq 0 $}{
        $r_{\mathcal{M}} \gets Fail$ 
    }
    
}
\Return $r_{\mathcal{M}}, \delta^{\mathcal{A}}$, $\delta^{\text{safe}}$
\caption{Algorithm for \oracle}
\label{algo:oracle}
\end{algorithm}
\subsubsection{Module-Specific Filter}\label{sec:filter}
The Module-Specific Filter aims to filter out less relevant module errors, as driving scenes farther from the termination have less impact on the final results~\cite{stocco2020misbehaviour, stocco2022thirdeye}.
The filter considers only module errors within a detection window \([T - \Delta t, T]\), where \( T \) is the timestamp of the occurrence of system failures in the scenario, and \( \Delta t \) is the detection window size. Therefore, the filtered module errors are calculated by:
\begin{equation}\label{eq:system_error}
    \hat{\delta}^{{M}} = \sum_{t=T-\Delta t}^{T} \mathbb{I}(\delta_t^{{M}} > \lambda^{{M}})
\end{equation}
where \( \lambda^{{M}} \) is the tolerance threshold for module ${M}$, and \( \mathbb{I} \) is an indicator function. The indicator function \( \mathbb{I}(condition) \) returns 1 if the \( condition \) is true and 0 otherwise. 

\subsubsection{Workflow of \oracle}
Algorithm~\ref{algo:oracle} illustrates the workflow of \oracle.
Specifically, the algorithm takes as inputs the scenario observation \( \mathcal{O}(s) \), the ADS \( \mathcal{A} \), and the user-specified module \( \mathcal{M} \), and it outputs three key results: the identification result \( r_{\mathcal{M}} \), the module errors's set \( \delta^{\mathcal{A}} \), the safety-critical distance \( \delta^{\text{safe}} \). 
Initially, \oracle creates an empty error set \( \delta^{\mathcal{A}} \) to store module errors and an identification flag \( r_{\mathcal{M}} \) set to `pass' (Line 1). Then, the algorithm calculates the safety-critical distance using Eq.~\ref{eq:safe} and checks for system failures (Lines 2-4), aiming to confirm the satisfaction of \textit{Definition~\ref{def-mccs}.a}.
Subsequently, the algorithm calculates and filters module errors for each module in the ADS \( \mathcal{A} \), storing these errors in \( \delta^{\mathcal{A}} \) for further analysis and feedback (Lines 5-8).
Once the user-specified module does not trigger errors (Line 9-10) or other modules do trigger errors (Line 11-12), the identification flag is set to `Fail' as they violate the requirements of \textit{Definition~\ref{def-mccs}.b} and \textit{Definition~\ref{def-mccs}.c}. 
Finally, \oracle returns the identification flag \( r_{\mathcal{M}} \), the module error set \( \delta^{\mathcal{A}} \), and the safety-critical distance \( \delta^{\text{safe}} \) (Line 13).

\subsection{Module-Specific Feedback} 

To provide guidance for searching {\mccs}s, we design a \feedback providing a feedback score, including two parts: (1) \textit{safety-critical score} and (2) \textit{module-directed score}. The \textit{safety-critical score} aims to guide the search for safety-critical scenarios that include system-level violations (i.e., collisions). To focus more specifically on the user-specified module, we introduce the \textit{module-directed score}, which provides guidance to bias the generation of safety-critical scenarios towards this module.

\subsubsection{Safety-critical Score}
We directly leverage the safety-critical distance from Section~\ref{sec:safe-metric} as our safety-critical feedback score, denoted by \( \phi_{s}^{\text{safe}} = \delta^{\text{safe}}_{s} \).
The safety-critical score \(\phi_{s}^{\text{safe}}\) quantifies the minimum distance between the ego vehicle and other objects over the time horizon \([0, T]\), capturing how close the vehicle comes to a collision scenario. A lower value of \(\phi_{s}^{\text{safe}}\) indicates a more dangerous situation for the ego vehicle.

\subsubsection{Module-directed Score} 
Given a user-specified module \( \mathcal{M} \), we calculate the \textit{module-directed score} as follows:
\begin{equation}
    \phi_{s}^{\mathcal{M}} = \sum_{t=T-\Delta t}^{T}\left( \delta_t^{\mathcal{M}} - \frac{\sum_{M \neq \mathcal{M}} \delta_t^{{M}}}{K - 1} \right)
\end{equation}
where \( K \) is the total number of modules in the ADS \(\mathcal{A}\), \( T \) is the termination timestamp of scenario \( s \), \( \Delta t \) is the detection window, and \( \delta_t^{{M}} \) represents the module-level errors for each module obtained from \oracle, as detailed in Section~\ref{sec:module-metric}. 
The first term, \( \delta_t^{\mathcal{M}} \), quantifies the errors specific to the user-specified module \( \mathcal{M} \). 
The second term, \( \frac{\sum_{M \neq \mathcal{M}} \delta_t^{{M}}}{K - 1} \), represents the average of the cumulative errors across all other modules. 
The score \( \phi_{s}^{\mathcal{M}} \) promotes the search for scenarios in which only the user-specified module \( \mathcal{M} \) exhibits errors during the detection window, while other modules do not. Therefore, this score can provide guidance to enhance the impact of the user-specified module \( \mathcal{M} \) on detected violations.

The final feedback score combines the \textit{safety-critical score} and the \textit{module-directed score} by:
\begin{equation}\label{eq:feedback}
    \phi_{s} =  \phi_{s}^{\mathcal{M}^k} - \phi_{s}^{\text{safe}}
\end{equation}
A larger \(\phi_{s}\) indicates that the scenario \( s \) is closer to becoming a {\mccs}. Consequently, \tool aims to generate {\mccs}s by maximizing this feedback score.

\begin{algorithm}[!t]
\small
\SetKwInOut{Input}{Input}
\SetKwInOut{Output}{Output}
\SetKwInOut{Para}{Parameters}
\SetKwProg{Fn}{Function}{:}{}
\SetKwFunction{AE}{\textbf{AdaptiveEvolver}}
\SetKwFunction{OI}{\textbf{ModuleCauseIdentifier}}
\SetKwComment{Comment}{\color{blue}// }{}
\Input{
Selected seed $s$ with feedback score ${\phi}_{s}$ \\
Maximum feedback score ${\phi}_{max}$ and Minimum feedback score ${\phi}_{min}$ in the seed corpus
}
\Output{
Mutated seed $s'$
}
$s' \gets s$ \Comment{Copy the selected scenario seed $s$}
$\lambda_{m} \gets \frac{{\phi}_{max} - {\phi}_{s}}{|{\phi}_{max} - {\phi}_{min}|}$ \Comment{Assign a dynamic threshold for determining mutation strategy}

\eIf{random() > $\lambda_{m}$}{
    \Comment{Fine-grained mutation: add small perturbation}
    $\mathbb{E}_{s'} \gets \mathbb{E}_{s'} + \text{GaussSample}(\lambda_m)$ \\
    \For{$P \in \mathbb{P}_{s'}$}{
    $W_{P}^{v} \gets W_{P}^{v} + \text{GaussSample}(\lambda_m)$ 
    }
}{
    \Comment{Coarse-grained mutation: regenerate new parameters}
    $\mathbb{E}_{s'} \gets \text{UniformSample}([\mathbb{E}_{min}, \mathbb{E}_{max}])$ \\
    \For{$P \in \mathbb{P}_{s'}$}{
    $W_{P}^{l} \gets \text{RouteGenerate}(\lambda_m)$ \\
    $W_{P}^{v} \gets \text{UniformSample}([V_{min}, V_{max}],\lambda_m)$
    }
}
\Return $s'$
\caption{Algorithm for \textit{Adaptive Mutation}}
\label{algo:mutation}
\end{algorithm}
\subsection{Adaptive Scenario Generation} 
To improve the searching performance of \tool, we design an \select mechanism including \textit{Adaptive Seed Selection} and \textit{Adaptive Mutation}, which adaptively generate new seed scenarios based on the feedback score obtained from \feedback.

\subsubsection{Adaptive Seed Selection} The seed selection process aims to choose a seed scenario from the seed corpus for further mutation, thereby generating a new scenario.
We design this selection process to favor seeds with higher feedback scores, indicating they are more likely to evolve into a {\mccs}.
Therefore, we assign a selection probability to each seed in the corpus based on the feedback score calculated by \feedback. The selection probability for each seed \( s \) is defined as:
\begin{equation}
    p(s) = \frac{\phi_{s} - \phi_{\min} + \epsilon}{\sum_{s' \in \mathbf{Q}} (\phi_{s'} - \phi_{\min} + \epsilon)}
\end{equation}
where \( \phi_{\min} \) is the global minimum feedback score in the corpus, \( \phi_{s} \) is the feedback score for seed \( s \), \( \epsilon \) is a small positive constant to ensure that the seed with the global minimum feedback score has a non-zero selection probability, and \( \mathbf{Q} \) denotes the set of all seeds in the corpus. This probability formulation ensures that seeds with higher feedback scores have a higher chance of being selected for mutation, thereby promoting the generation of scenarios with a higher likelihood of evolving into a {\mccs}.

\subsubsection{Adaptive Mutation} Beyond seed selection, we also design an adaptive mutation strategy that applies different mutation methods based on the feedback score of each seed. 
Algorithm~\ref{algo:mutation} presents the details of \textit{Adaptive Mutation}. 
Specifically, given a scenario \( s \), the \textit{Adaptive Mutator} first copies the source seed \( s \) to \( s' \) (Line 1) and calculates a dynamic threshold \( \lambda_m = \frac{\phi_{\text{max}} - \phi_s}{|\phi_{\text{max}} - \phi_{\text{min}}|} \in [0, 1] \) by normalizing the feedback score \( \phi_s \), ensuring \( \lambda_m \) (Line 2).
Then, the mutation selects either \textit{fine-grained mutation} or \textit{coarse-grained mutation} based on the derived dynamic threshold $\lambda_m$ (Line 3-11). Seeds with higher feedback scores (resulting in smaller dynamic thresholds) are regarded as closer to {\mccs}s; therefore, we employ \textit{fine-grained mutation} to add Gaussian noise to the weather parameters (Line 4) and the speeds of each object (Lines 5-6). Otherwise, for seeds with lower feedback scores (resulting in larger dynamic thresholds), we utilize \textit{coarse-grained mutation} to introduce more significant variations. These include changing the trajectory waypoints of each object and altering environmental conditions through uniform sampling (Lines 7-11).
Finally, a mutated scenario \( s' \) is produced (Line 12).

\section{Evaluation}\label{sec: Evaluation}
In this section, we aim to empirically evaluate the effectiveness of \(\text{\tool}\) in generating {\mccs}s. Specifically, we will address the following research questions:

\newcommand{\rqone}{Can \(\text{\tool}\) effectively generate {\mccs}s across different modules of the ADS?}
\newcommand{\rqtwo}{How accurate is \(\text{\oracle}\) in identifying {\mccs}s?}
\newcommand{\rqthree}{What is the usefulness of \(\text{\feedback}\) and \(\text{\select}\) in \(\text{\tool}\)?}
\newcommand{\rqfour}{How does \tool perform from the perspective of testing efficiency?}
\noindent \textbf{RQ1: } \rqone{}

\noindent \textbf{RQ2: } \rqtwo{}

\noindent \textbf{RQ3: } \rqthree{}

\noindent \textbf{RQ4: } \rqfour{}

To address these research questions, we conducted experiments using the following settings:

\textbf{Environment.} We conducted our experiments using Pylot~\cite{gog2021pylot} and CARLA~\cite{dosovitskiy2017carla}. Pylot, a widely popular open-source, multi-module ADS platform, includes modules for Perception $\mathcal{M}^{\text{perc}}$, Prediction $\mathcal{M}^{\text{pred}}$, Planning $\mathcal{M}^{\text{plan}}$, and Control $\mathcal{M}^{\text{ctrl}}$. CARLA is a high-fidelity simulator that is compatible with Pylot.

\textbf{Driving Scenarios.} 
We evaluate \tool on four representative scenarios derived from the NHTSA pre-crash typology~\cite{najm2007pre}, which is also widely utilized in existing ADS testing techniques~\cite{li2020av, cheng2023behavexplor}. Specifically, these scenarios are:

\noindent \textit{S1:} The ego vehicle starts in a weave zone and merges onto the highway.

\noindent \textit{S2:} The ego vehicle leaves the highway via an exit ramp.

\noindent \textit{S3:} The ego vehicle turns left at an uncontrolled intersection.

\noindent \textit{S4:} The ego vehicle turns right at an uncontrolled intersection; a full stop is required before turning.

\textbf{Baselines.}
We selected three state-of-the-art system-level testing methods for the comparisons. We first select a \textit{Random} method that randomly generates scenarios. Additionally, we included two representative state-of-the-art ADS testing techniques for reference: \textit{AVFuzzer}~\cite{li2020av} and \textit{BehAVExplor}~\cite{cheng2023behavexplor}. \textit{AVFuzzer} aims to efficiently identify collisions, whereas \textit{BehAVExplor} focuses on discovering a more diverse set of safety-critical scenarios. 
Note that \textit{AVFuzzer} and \textit{BehAVExplor} were originally evaluated using the Apollo~\cite{baiduapollo} with LGSVL~\cite{rong2020lgsvl}. However, since LGSVL was sunsetted in 2022~\cite{LGSVLSunsetting}, we adapted these techniques to our simulation environment, Pylot with CARLA, for comparison. 

\textbf{Metrics.} To evaluate the effectiveness of \tool, we utilize the metric $\#\mathcal{M}^{k}$, which quantifies the number of generated \mccs\ for the user-specific module $\mathcal{M}^{k}$. 
Additionally, we aim to confirm that the detected {\mccs}s are truly induced solely by errors in the module \( \mathcal{M} \). To this end, we define a repair rate metric as \( \%\mathcal{M}^{k} = \frac{\#\mathcal{M}^{k}_{\text{rep}}}{\#\mathcal{M}^{k}} \times 100\% \) to verify the correctness of \oracle, where \( \#\mathcal{M}^{k}_{\text{rep}} \) denotes the number of \mccs\ successfully repaired.
Here, $k$ is the identifier for the module $\mathcal{M}^{k}$; for instance, $k = \text{perc}$ denotes the Perception module. 
Specifically, we rerun and repair the detected \mccs\ by replacing the outputs of $\mathcal{M}^{k}$ with the ground truth. Subsequently, we count the number of \mccs\ that have been repaired and no longer present any safety-critical issues.

\textbf{Implementation.} 
According to our preliminary study (see Section \ref{sec: perliminary_study}), we set the tolerance thresholds $\lambda_{\mathcal{M}^k}$ for Perception, Prediction, and Control modules as $\lambda_{\mathcal{M}^{\text{perc}}} = 0.5$, $\lambda_{\mathcal{M}^{\text{pred}}} = 0.1$, $\lambda_{\mathcal{M}^{\text{plan}}} = 0$, and $\lambda_{\mathcal{M}^{\text{ctrl}}} = 0.05$, respectively. The detection window size $\Delta t$ is set to 0.5 seconds because this is the smaller length of the prediction and planning module output in Pylot's default configuration.
In our experiments, we utilize the synchronized mode in CARLA to mitigate the influence of non-determinism inherent in the simulation-based execution of the ADS. We repeat each experiment three times with different random seeds to report the average of the results. For each run, we use a consistent budget of three hours, which we found sufficient for comparison. 

\begin{table}[!t]
    \centering
    \caption{Comparison results with baselines.}
    \vspace{-10pt}
    \resizebox{\linewidth}{!}{
    \begin{tabular}{l|ccccc|ccccc|ccccc|ccccc}
    \toprule
         \multirow{2.5}*{Method} & \multicolumn{5}{c|}{\#$\mathcal{M}^{\text{perc}}$} & \multicolumn{5}{c|}{\#$\mathcal{M}^{\text{pred}}$} & \multicolumn{5}{c|}{\#$\mathcal{M}^{\text{plan}}$} & \multicolumn{5}{c}{\#$\mathcal{M}^{\text{ctrl}}$}\\
        \cmidrule(lr){2-6}\cmidrule(lr){7-11}\cmidrule(lr){12-16}\cmidrule(lr){17-21}
          & \textit{S1} & \textit{S2} & \textit{S3} & \textit{S4} & \cellcolor{lightgray!20}\textit{Sum.} & \textit{S1} & \textit{S2} & \textit{S3} & \textit{S4} & \cellcolor{lightgray!20}\textit{Sum.} & \textit{S1} & \textit{S2} & \textit{S3} & \textit{S4} & \cellcolor{lightgray!20}\textit{Sum.} & \textit{S1} & \textit{S2} & \textit{S3} & \textit{S4} & \cellcolor{lightgray!20}\textit{Sum.} \\
         \midrule
         \textit{Random} & 2.3 & 1.7 & 2.0 & 1.7 & \cellcolor{lightgray!20}7.7 & 6.3 & 11.3 & 4.0 & 1.7 & \cellcolor{lightgray!20}23.3 & 11.3 & 11.7& 7.3 & 3.7& \cellcolor{lightgray!20}34.0 & 3.3 & 0 & 1.3 & 0 & \cellcolor{lightgray!20}4.6\\
         \textit{AVFuzzer} & 3.3 & 4.7 & 1.3 & 3.7 & \cellcolor{lightgray!20}13.0 & 11.7 & 6.7 & 3.7& 3.3 & \cellcolor{lightgray!20}25.3 & 7.7 & 10.7 & 5.7 & 5.3 & \cellcolor{lightgray!20}29.3 & 3.7 & 0.7 & \textbf{2.0} & 0.3 & \cellcolor{lightgray!20}6.7\\
         \textit{BehAVExplor} & 5.3 & 4.3& 4.3 & 5.7 & \cellcolor{lightgray!20}18.3 & 6.7 & 10.3 & 6.0 & 1.3 & \cellcolor{lightgray!20}19.3 & 11.0 & 13.3 & 7.3 & 2.7& \cellcolor{lightgray!20}34.3 & 4.7 & 1.3 & 1.0 & 0.0 & \cellcolor{lightgray!20}7.0\\
         \tool & \textbf{23.0} & \textbf{7.7} & \textbf{14.0} & \textbf{10.7} & \cellcolor{lightgray!20}\textbf{55.3} & \textbf{28.7} & \textbf{24.3} & \textbf{13.3} & \textbf{9.0} & \cellcolor{lightgray!20}\textbf{75.3} & \textbf{26.0} & \textbf{21.0} & \textbf{18.0} & \textbf{6.7} & \cellcolor{lightgray!20}\textbf{71.7} & \textbf{6.0} & \textbf{5.0} & \textbf{2.0} & \textbf{1.3}&\cellcolor{lightgray!20} \textbf{14.3}\\
         \bottomrule
    \end{tabular}
    }
    
    \label{tab:RQ1}
\end{table}

\subsection{RQ1: Effectiveness of \tool}\label{sec:exp_rq1}

\subsubsection{Comparative Results} 
Table~\ref{tab:RQ1} presents the comparison results on the number of generated \mccs\ $\#\mathcal{M}^{k}$, for each individual module in the ADS, covering Perception $\#\mathcal{M}^{\text{perc}}$, Prediction $\#\mathcal{M}^{\text{pred}}$, Planning $\#\mathcal{M}^{\text{plan}}$, and Control $\#\mathcal{M}^{\text{ctrl}}$. 
From the results, we can find that \tool achieves significantly better performance than other baselines in generating \mccs\ for each module. 
Specifically, in terms of the \textit{Sum.} numbers of $\#\mathcal{M}^{k}$, \tool substantially outperforms the best baseline: \textit{BehAVExplor} on $\#\mathcal{M}^{\text{perc}}$ (55.3 vs. 18.3), \textit{AVFuzzer} on $\#\mathcal{M}^{\text{pred}}$ (75.3 vs. 25.3), \textit{BehAVExplor} on $\#\mathcal{M}^{\text{plan}}$ (71.7 vs. 34.3), and \textit{BehAVExplor} on $\#\mathcal{M}^{\text{ctrl}}$ (6.0 vs. 4.7). 
It is worth noting that \textit{BehAVExplor} outperforms other baselines in most modules due to its diversity feedback mechanism. This mechanism encourages the ego vehicle to explore different behaviors, making it more likely to generate safety-critical scenarios caused by different modules. 
By comparing results across different modules, we find that \tool identifies \#\mccs\ in descending order: Prediction (75.3) $>$ Planning (71.7) $>$ Perception (55.3) $>$ Control (14.3). This hierarchy indicates that in the ADS, the robustness of the Prediction, Planning, and Perception modules is comparatively lower and requires further development. 
We note that both comparison methods and \tool only detects a very small number of $\#\mathcal{M}^{\text{ctrl}}$. 
In the experiment, the control module is based on PID~\cite{johnson2005pid}, and with appropriate parameter settings, PID controllers generally exhibit high robustness~\cite{aastrom1993automatic}. On the other hand, scenario-based testing cannot directly generate sudden load changes and other variations commonly encountered in PID testing~\cite{brannstrom2010model}. Nonetheless, \tool still achieves an improvement over baseline methods in detecting $\mathcal{M}^{\text{Ctrl}}CCS$.
Moreover, \tool consistently achieves better performance than all baselines across the four scenarios, \textit{S1} to \textit{S4}, demonstrating its generalization capability in various situations. 

\begin{figure}[!t]
    \centering
    \subfloat[\scriptsize $\mathcal{M}^{\text{perc}}$ICS]{\includegraphics[width=0.22\linewidth]{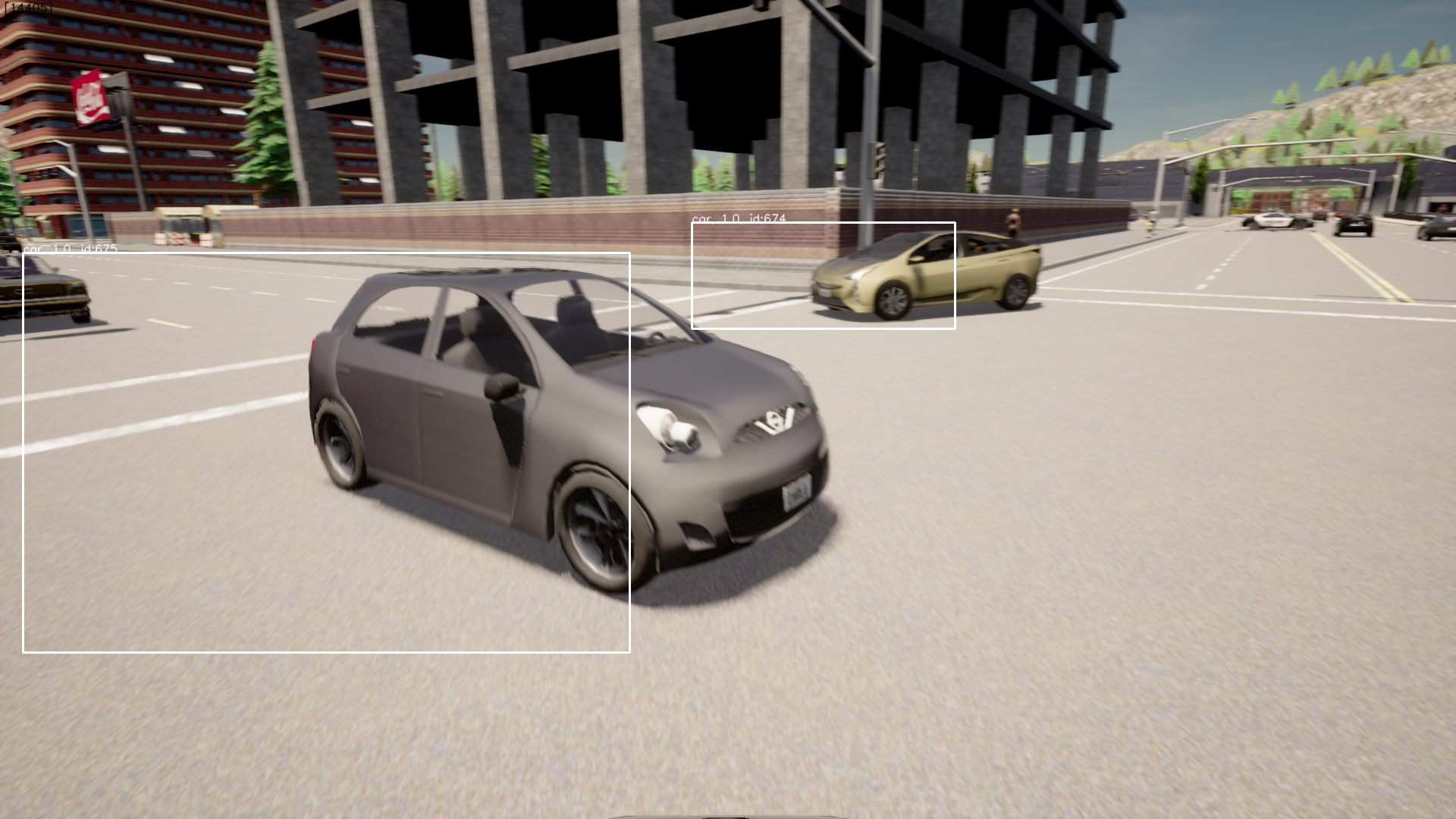}}
    \quad
    \subfloat[\scriptsize $\mathcal{M}^{\text{pred}}$ICS]{\includegraphics[width=0.22\linewidth]{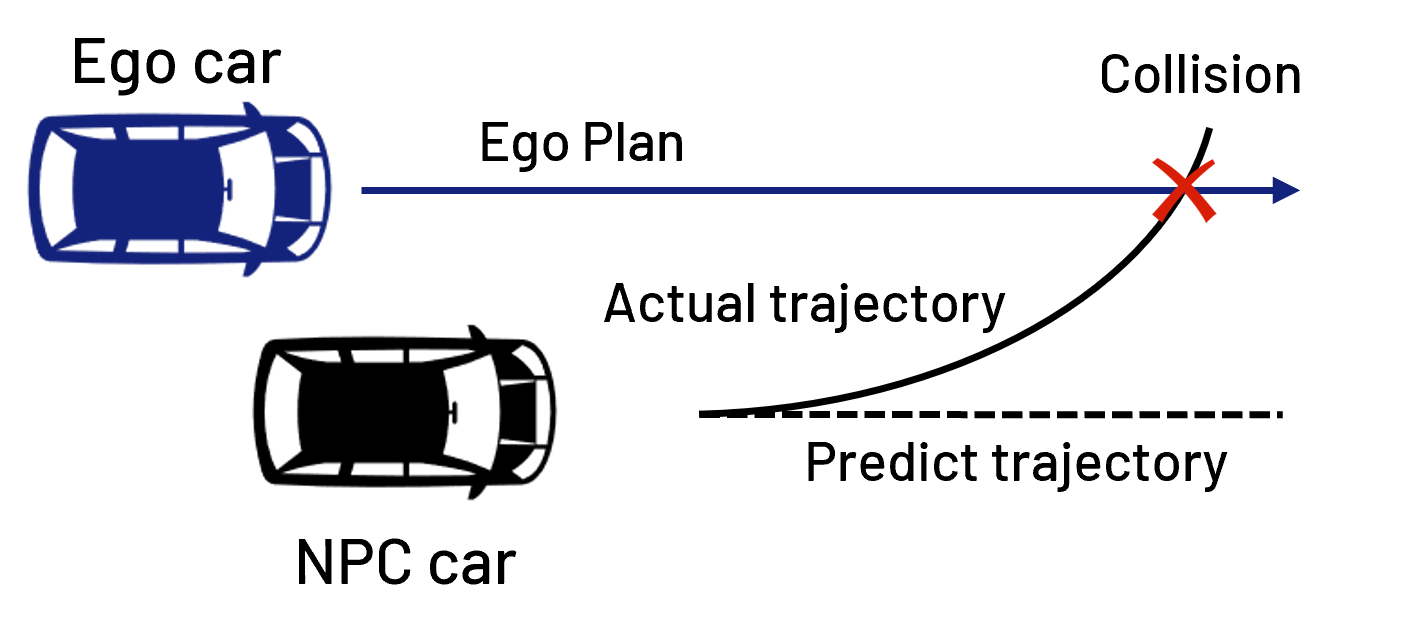}}
    \quad
    \subfloat[\scriptsize $\mathcal{M}^{\text{plan}}$ICS]{\includegraphics[width=0.22\linewidth]{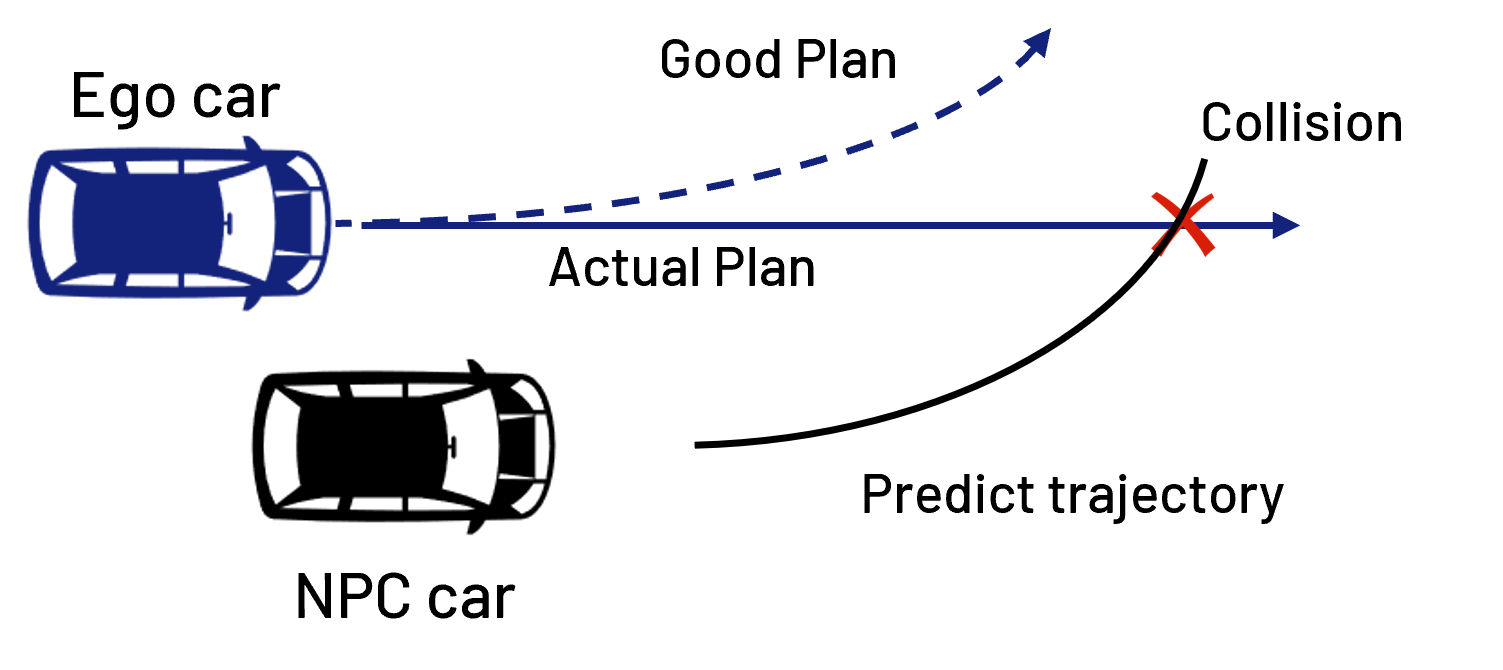}}
    \quad
    \subfloat[\scriptsize $\mathcal{M}^{\text{ctrl}}$ICS]{\includegraphics[width=0.22\linewidth]{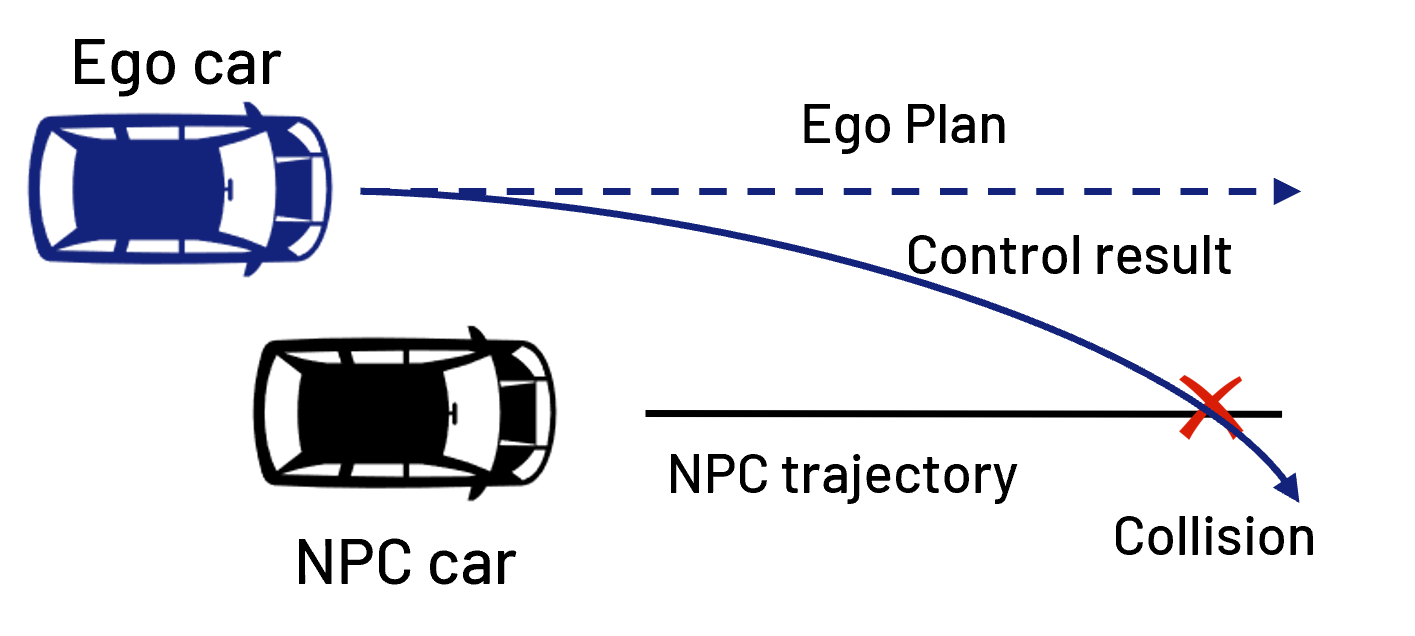}}
    \caption{Cases of {\mccs}s detected by \tool for different specific modules.}
    \label{fig:case}
\end{figure}
\subsubsection{Case Study} Fig.~\ref{fig:case} presents four representative \mccs\ examples for modules in the ADS, respectively. 

(a) $\mathcal{M}^{\text{perc}}$ICS. Fig.~\ref{fig:case}(a) illustrates a collision caused solely by the Perception module. In this scenario, the ego car turns left and erroneously detects the NPC cars' locations as they shift to the left. As a result, the ADS assumes a safe distance between the detected NPC cars and, therefore, maintains its acceleration decision, ultimately leading to a collision.

(b) $\mathcal{M}^{\text{pred}}$ICS. Fig.~\ref{fig:case}(b) illustrates a collision scenario resulting from inaccurate predictions of surrounding NPC cars' movements by the Prediction module. In this instance, the ego car maintains its lane, while a nearby NPC car on the right initiates a lane change to cut in. However, the Prediction module erroneously predicts that the NPC car will stay in its current lane. Consequently, the ADS continues its original trajectory without accounting for the NPC car's lane change, which ultimately leads to a collision.

(c) $\mathcal{M}^{\text{plan}}$ICS. Fig.~\ref{fig:case}(c) illustrates a collision caused by the Planning module of the ego car making an unsafe decision. In this case, the ego car proceeds along its initially planned path (Actual Plan), even though a safer alternative trajectory (Good Plan) is available. The ego car fails to account for the NPC car's predicted trajectory, which leads to a close interaction. By not choosing the safer path, the Planning module's decision ultimately results in a collision with the NPC car.

(d) $\mathcal{M}^{\text{ctrl}}$ICS. Fig~\ref{fig:case}(d) is a collision caused by the Control module in the ADS. Specifically, the ego of the upstream Planning makes a safe planning path (Ego Plan), while the Control module can not adjust the ego-motion according to the Ego Plan in time. The latency response in the Control module finally results in a collision with the NPC car. 

\begin{ansbox}
\textbf{Answer to RQ1:} \tool can effectively generate {\mccs}s for user-specified module $\mathcal{M}^{k}$ in the ADS, covering Perception, Prediction, Planning and Control modules.  
\end{ansbox}

\begin{table}[!t]
    \centering
    \caption{Correctness of \oracle for the Perception and Prediction modules.}
    \vspace{-10pt}
    \resizebox{\linewidth}{!}{
        \begin{tabular}{c|ccc|ccc|ccc|ccc}
        \toprule
         \multirow{2.5}*{Scenario} & \multicolumn{3}{c|}{Perception} & \multicolumn{3}{c|}{Prediction} & \multicolumn{3}{c|}{Planning} & \multicolumn{3}{c}{Control}\\
         \cmidrule(lr){2-4}\cmidrule(lr){5-7}\cmidrule(lr){8-10}\cmidrule(lr){11-13}
         & \#$\mathcal{M}^{\text{perc}}$ & \#$\mathcal{M}^{\text{perc}}_{\text{rep}}$ & \%$\mathcal{M}^{\text{perc}}$ & \#$\mathcal{M}^{\text{pred}}$ & \#$\mathcal{M}^{\text{pred}}_{\text{rep}}$ & \%$\mathcal{M}^{\text{pred}}$ &
         \#$\mathcal{M}^{\text{plan}}$ & \#$\mathcal{M}^{\text{plan}}_{\text{rep}}$ & \%$\mathcal{M}^{\text{plan}}$  &
         \#$\mathcal{M}^{\text{ctrl}}$ & \#$\mathcal{M}^{\text{ctrl}}_{\text{rep}}$ & \%$\mathcal{M}^{\text{ctrl}}$ \\
         \midrule
         \textit{S1} & 23.0 & 21.0 & 91.3\% & 28.7 & 24.3 &85.0\% & 26.0 & 26.0 & 100.0\% & 6.0& 6.0 & 100.0\%\\
         \textit{S2} & 7.7 & 7.3 & 96.1\% &  24.3 & 22.0 & 90.5\% & 21.0& 21.0 & 100.0\% & 5.0& 5.0 & 100.0\%\\
         \textit{S3} & 14.0 & 11.7 &83.6\% & 13.3 & 12.7 & 95.5\% & 18.0& 18.0 & 100.0\%& 2.0& 2.0 & 100.0\%\\
         \textit{S4} & 10.7 & 9.3 &87.7\% & 9.0 & 8.3 & 92.2\% & 6.7& 6.7 & 100.0\% &1.3 & 1.3 & 100.0\%\\
         \textit{Sum.} &55.3 & 49.3 & 89.2\% &75.3 & 67.3 & 89.3\% & 71.7& 71.7 & 100.0\% &14.3 & 14.3 & 100.0\%\\
         \bottomrule
    \end{tabular}
    }
    \vspace{-10pt}
    \label{tab:rq2}
\end{table}
\subsection{RQ2: Correctness of \oracle}
\label{sec:exp_rq2}

\subsubsection{Setup}
To evaluate the fidelity of \oracle, we rerun all detected {\mccs}s by substituting the outputs of module $\mathcal{M}^{k}$ with perfect outputs starting from timestamp $t_{\mathcal{M}^{k}}$. 
Here, \( t_{\mathcal{M}^{k}} \) denotes the timestamp of the first error in module \( \mathcal{M}^{k} \) detected within the detection window (defined in Section~\ref{sec:filter}).
For the Perception module, we replace its results with bounding boxes retrieved from CARLA, ensuring that the module's outputs align with the ground truth. For the Prediction module, we obtain ground truth data from the collected dataset. Specifically, when rerunning a scenario $s$ with its collected observations $\mathcal{O}(s)$, at time $t$, we retrieve the actual trajectory of the NPC object over the interval $[t, t + H_{\text{pred}}]$ from the simulator's observation $\mathcal{Y}(s) \in \mathcal{O}(s)$ and use this as the ground truth for the Prediction module.
Considering the vast space of possible planned trajectories and control commands, we use the safest solution, which involves immediate braking, as the ground truth. This approach ensures that both the planning module and the control module consistently make the safest possible decisions.

\subsubsection{Results} 
Table~\ref{tab:rq2} presents the results for the correctness of \oracle. By replacing the module outputs with a safe ground truth, we observe that most of the detected {\mccs}s can be repaired, with average repair rates of 89.2\%, 89.3\%, 100\% and 100\% for the Perception, Prediction, Planning, and Control modules, respectively. These results demonstrate that our proposed \oracle can accurately identify the module-level root causes.
Notably, the repair rates for the Perception and Prediction modules are not 100\% (89.2\% and 89.3\%, respectively). This is primarily because these two modules are upstream; even when corrected with safe ground truth, it does not necessarily ensure that the downstream modules will continue to make safe decisions.

\begin{ansbox}
    \textbf{Answer to RQ2:} \oracle can accurately identify the module whose errors are the root cause of safety-critical violations.
\end{ansbox}

\begin{table}[!t]
\caption{Results of ablation study on Prediction module.}
\vspace{-10pt}
    \centering
    \resizebox{0.8\linewidth}{!}{
    \begin{tabular}{c|cccccc}
        \toprule
        Scenario & \tool & \textit{w/o Fine} & \textit{w/o Coarse} & \textit{w/o Select} & \textit{w/o F-$\mathcal{M}$}  & \textit{Random} \\
        \midrule
        \textit{S1} & 28.7 & 14.3 & 7.7 &  15.3&7.7& 6.3 \\
        \textit{S2} & 24.3 & 17.0 & 13.3 &  20.7 & 15.7&11.3 \\
        \textit{S3} & 13.3 & 7.7 & 5.0 & 9.3 &6.7 & 4.0 \\
        \textit{S4} & 9.0 & 2.7 & 1.3 & 7.7 & 3.0 & 1.7\\
        \midrule
        \textit{Sum.} & 75.3 & 39.7 & 27.3 & 53 & 33 & 23.3 \\
        \bottomrule
    \end{tabular}}
    \vspace{-10pt}
\label{table:RQ3}
\end{table}
\subsection{RQ3: Usefulness of \feedback and \select}
We assess the usefulness of key components in the fuzzing process in \tool, namely \feedback and \select. To achieve this, we conducted a thorough evaluation by configuring a series of variants of \tool and then proceeded to evaluate their effectiveness. Currently, we verify this on the Prediction module.

\subsubsection{\feedback} To verify the effectiveness of feedback, we compare \tool with two variants: (1) \textit{w/o F-$\mathcal{M}$}, which removes the module-directed feedback from Eq.~\ref{eq:feedback} to assess the impact of module-directed score on detecting {\mccs}s; (2) \textit{Random}, which serves as a reference by reflecting the influence of the entire feedback mechanism on detecting {\mccs}s.
As shown in Table~\ref{table:RQ3}, we find that \textit{Random} generates the fewest {\mccs}s (23.3), underscoring the importance of the designed feedback. Comparing \textit{w/o F-$\mathcal{M}$} with \tool, we observe that \tool generates more {\mccs}s than \textit{w/o F-$\mathcal{M}$} (75.3 vs. 33), highlighting the value of module-directed feedback in detecting a greater number of {\mccs}s.

\subsubsection{\select} For \select, we design three variants: (1) \textit{w/o Fine}, which removes the fine-grained mutation from the adaptive mutation to evaluate the effectiveness of this mutation level; (2) \textit{w/o Coarse}, similar to \textit{w/o Fine}, but removes the coarse-grained mutation; and (3) \textit{w/o Select}, which changes the adaptive seed selection to random seed selection to assess the usefulness of the adaptive mechanism. 
From the results shown in Table~\ref{table:RQ3}, we find that removing either the fine-grained mutation or the coarse-grained mutation reduces the number of detected {\mccs}s, with \textit{w/o Fine} detecting only 41.7 {\mccs}s and \textit{w/o Coarse} detecting 27.3 {\mccs}s. This demonstrates the effectiveness of our proposed adaptive mutation.
By comparing \textit{w/o Select} with \tool, we observe that \textit{w/o Select} detects only 53 {\mccs}s, indicating that adaptive seed selection is an effective strategy for generating {\mccs}s.

\begin{ansbox}
\textbf{Answer to RQ3:} Both \feedback and \select are useful for \tool in detecting {\mccs}s effectively.
\end{ansbox}

\subsection{RQ4: Performance of \tool}
\begin{table}[h]
    \centering
    \caption{Results of time performance in seconds (s). * denotes a negligible minimal value.}
    \vspace{-5pt}
   \resizebox{0.75\linewidth}{!}{
    \begin{tabular}{l|cccc|c}
    \toprule
    Method & Seed Selection & Mutation & Feedback & Simulation & Total\\
    \midrule
         Random& 0.01* & 1.12 & N/A & 213.32 & 214.44 \\
         AVFuzzer& 0.01* & 1.99& 0.01* & 219.58 & 221.48\\
         BehAVExplor & 0.01* & 1.64 & 1.09 & 212.03 & 214.76\\
         \midrule
         \tool & 0.01* & 3.23 & 0.11 &172.79 & 176.13\\
         \bottomrule
    \end{tabular}
   }
   \vspace{-10pt}
    \label{tab:performace}
\end{table}

We further assess the time performance of different components in \tool, including the overhead of Seed Selection, Mutation, Feedback, and Simulation. Specifically, we analyze the average time required to process a scenario. Table~\ref{tab:performace} shows the overall results for all tools. From the results, we find that the simulation process occupies the majority of the time. For instance, on average, \tool takes 176.13 seconds to process a scenario, with 172.79 seconds (98.1\%) spent running the scenario in the simulator. All tools spend negligible time on seed selection, as it is a simple sampling operation. For mutation, \tool requires more time than others due to the additional overhead of dynamic mutation. Our feedback calculation takes an average of 0.11 seconds, which is faster than \textit{BehAVExplor}. In summary, the computation time of \tool remains within an acceptable range.

\begin{ansbox}
\textbf{Answer to RQ4:} \tool demonstrates efficiency, with the majority of the time (98.1\%) spent in the simulation phase, while the main algorithmic components consume approximately 3.34 seconds (1.9\%) of the total processing time.
\end{ansbox}

\section{Discussion}
This paper is the first to investigate ADS failures from module-level root causes. Although \tool can effectively and efficiently detect {\mccs}s, there are several potential areas for improvement that warrant further discussion.

\noindent \textbf{Non-\mccs Failures}
In \tool, we employ a strict oracle to determine \mccs by requiring that only one module fails before a collision occurs. Consequently, many failure scenarios are not classified as being caused by a specific module, rendering them non-\mccs failures. However, in some non-\mccs scenarios, if the ground truth for a specific module is provided, as demonstrated in RQ2, normal operation might resume. These potential \mccs may warrant further investigation.

\noindent \textbf{Future Works} There are two potential directions for future work. (1) Currently, \tool only considers single-module analysis. However, our \tool can be easily extended to support the detection of {\mccs}s induced by multiple modules, which we plan to explore in future work. (2) In our individual module metrics, we focus solely on safety as the metric. Nevertheless, evaluating the ADS from non-safety-critical aspects is also important. In future work, we will incorporate additional metrics to broaden the evaluation scope.

\section{Threats to Validity}\label{sec: threats}

\noindent\textbf{Internal Validity.}
The accuracy of Root Cause Analysis is critical for \abb{}, as it affects the evaluation of the generated test scenarios and serves as the foundation for the subsequent fuzzing process. To achieve the most precise detecting for \mccs, we employ the strictest criterion: within the collision's effect window, there must be one and only one module that experiences an error for us to attribute the collision to that module's fault. Although this approach may not always result in a relatively high proportion of qualifying scenarios(see Section~\ref{sec:exp_rq1}), it ensures that the collisions in the generated test scenarios are indeed associated with specific modules(see Section~\ref{sec:exp_rq2}).

\noindent \textbf{External Validity.}
Since different ADSs employ varying combinations of modules and each module implementation has its own strengths and weaknesses, the experimental results can only be fully guaranteed to be directly related to the specific ADS and its corresponding model under test. To address this limitation, we will subsequently conduct tests on different ADSs and replace the existing module implementations within the ADS to further explore and validate our approach.

\section{Related Works}\label{sec: RelatedWorks}

\noindent  \textbf{Root Cause Analysis for AI Systems}
Inspired by testing approaches in other AI systems\cite{shi2024finding, xie2023mosaic, wang2022exploratory, yu2024survey, bothe2020neuromorphic,kim2020control, hossen2023care}, recent years have seen attempts to introduce root cause analysis into the testing of ADS and related robotic systems.
Swarmbug\cite{jung2021swarmbug} treats the AI system as a black box, and proposes \textit{Degree of Causal Contribution} to measure how configurations affect the behaviour of the swarm robotics. In contrast, the \oracle in \tool focuses more on the outputs of individual modules within the system, allowing us to identify the root causes of issues at a finer granularity.
RVPLAYER\cite{choi2022rvplayer} and ROCAS\cite{feng2024rocas} propose an algorithm that replays the accident scenarios and checks if the accident can be avoided by changing some parameters to locate the root cause. 
Compared to the methods in these two works, our \oracle establishes analysis metrics for each module and develops a module-to-system mapper to address the challenge of module-level error measurement. This allows for the analysis to be conducted with just a single run of the accident scenario, making the process more efficient and accurate in identifying the \mccs.

\noindent \textbf{Search-based Scenario Generation for ADS Testing}
Search-based methods have become one of the most popular algorithms in scenario-based testing due to their ability to efficiently explore complex scenario spaces\cite{ding2023survey,zhong2021survey}. From an algorithmic framework perspective, it can be categorized into evolutionary algorithm\cite{han2021preliminary,tang2021systematic,zhou2023specification,calo2020generating,humeniuk2022search}, model-based searching\cite{haq2022efficient,haq2023many,zhong2022neural,feng2023dense,li2023generative}, and fuzzing methods\cite{pang2022mdpfuzz,cheng2023behavexplor,fu2024icsfuzz,li2020av,cheng2024evaluating,cheng2024drivetester} which we use in \tool. 
Unlike these methods, \tool goes beyond efficiently generating collision or failure scenarios. It establishes a relationship between module errors and system failures, generating \mccs for specific modules within the ADS. This approach is designed to better help ADS developers in enhancing their system components.
\section{Conclusion}\label{sec: Conclusion}

In this work, we introduce \abb{}, a fuzzing method to generate specific-module-sensitive test scenarios for ADS testing.
\abb{} first build a root analysis function to locate which modules are corresponded to the collision. Building upon this function, \abb{} calculates root cause scores and customizes the seed selection and mutation modules accordingly. 
Extensive experiments are conducted with various primary scenarios, specific modules and fuzzing methods to validate the effectiveness of \abb{}. Experimental results demonstrate that \abb{} is highly effective in generating collision scenarios caused by errors in specific modules.

\bibliographystyle{ACM-Reference-Format}
\bibliography{acmart}
\end{document}